\documentclass[floatfix,%
 aip,
 amsmath,amssymb,
 reprint
,
]{revtex4-2}

\usepackage{graphicx}
\usepackage{dcolumn}
\usepackage{bm}
\usepackage[utf8]{inputenc}
\usepackage[T1]{fontenc}
\usepackage{mathptmx}
\usepackage{etoolbox}
\usepackage{gensymb}
\usepackage{comment}

\makeatletter
\def\@email#1#2{%
 \endgroup
 \patchcmd{\titleblock@produce}
  {\frontmatter@RRAPformat}
  {\frontmatter@RRAPformat{\produce@RRAP{*#1\href{mailto:#2}{#2}}}\frontmatter@RRAPformat}
  {}{}
}%
\makeatother
\begin{document}

\preprint{AIP/123-QED}

\title{Ionomeric extracellular matrices for dynamic soft robotic tissue engineering devices through protein sulfonation}
\author{Matthew K. Burgess}
\affiliation{ 
Institute of Biomedical Engineering, Department of Engineering Science, University of Oxford, Marcela Botnar Wing, Windmill Road, Oxford OX3 7LD, United Kingdom}
\author{Ryan T. Murray}
\affiliation{ 
Institute of Biomedical Engineering, Department of Engineering Science, University of Oxford, Marcela Botnar Wing, Windmill Road, Oxford OX3 7LD, United Kingdom}
\author{Veronica M. Lucian}
\affiliation{ 
Institute of Biomedical Engineering, Department of Engineering Science, University of Oxford, Marcela Botnar Wing, Windmill Road, Oxford OX3 7LD, United Kingdom}
\author{Zekun Liu}
\affiliation{ 
Institute of Biomedical Engineering, Department of Engineering Science, University of Oxford, Marcela Botnar Wing, Windmill Road, Oxford OX3 7LD, United Kingdom}
\author{Robin O. Cleveland}
\affiliation{ 
Institute of Biomedical Engineering, Department of Engineering Science, University of Oxford, Marcela Botnar Wing, Windmill Road, Oxford OX3 7LD, United Kingdom}
\author{Callum J. Beeston}
\affiliation{Department of Materials Science and Metallurgy, University of Cambridge, 27 Charles Babbage Road, CB3 0FS, United Kingdom}
\author{Malavika Nair*}%
\email{malavika.nair@eng.ox.ac.uk}
\affiliation{ 
Institute of Biomedical Engineering, Department of Engineering Science, University of Oxford, Marcela Botnar Wing, Windmill Road, Oxford OX3 7LD, United Kingdom}
\affiliation{Department of Materials Science and Metallurgy, University of Cambridge, 27 Charles Babbage Road, CB3 0FS, United Kingdom}
%

\date{\today}

\begin{abstract}
Conventional tissue engineering methodologies frequently depend on pharmacological strategies to induce or expedite tissue repair. However, bioengineered strategies incorporating biophysical stimulation have emerged as promising alternatives. Electroactive materials facilitate the provision of controlled electrical, mechanical, and electromechanical stimuli, which support cell proliferation and tissue remodelling. Despite their ability to supply external electrical and mechanical stimuli to the tissue microenvironment, the electroactive polymers in use today often lack critical biochemical signals essential for native-like cell-cell and cell-scaffold interactions, thereby constraining their regenerative capabilities. To address the demand for biomimetic materials that possess enhanced capabilities in promoting cell and tissue stimulation, we present the development of a novel class of polymers called ionomeric extracellular matrices (iECMs). By utilising the linker-mediated conjugation of sulfonic acid biomolecules (taurine) to the backbone of an extracellular matrix protein (collagen), we illustrate the potential of iECMs as the first electromechanical actuating material platform derived entirely from ECM materials, paving the way for dynamic and soft-robotic platforms for a wide range of tissue engineering applications.
\end{abstract}

\maketitle 
\onecolumngrid
\section{Main}

The restoration of functional capabilities such as sensation and movement in damaged tissues remains a substantial challenge in the field of tissue engineering. The advancement of highly tissue-compatible materials and scaffold architectures has significantly enhanced the rate of repair, demonstrating the efficacy of static structures as platforms for regeneration \cite{bettingerEngineeringSubstrateTopography2009a, výbornýGenipinEDCCrosslinking2019, basurtoAlignedElectricallyConductive2021, heimbachMulticenterPostapprovalClinical2003, salvatoreMimickingHierarchicalOrganization2021, baklaushevTissueEngineeredNeural2019}. Nevertheless, tissues are inherently dynamic, continually subjected to endogenous and exogenous physical and chemical stimuli that influence cellular processes \cite{caoPhysicalCuesScaffolds2024, kougiasArterialBaroreceptorsManagement2010, moeendarbaryCellMechanicsPrinciples2014, nomuraMolecularEventsCaused2000}. While research efforts have predominantly concentrated on the utilisation of growth factors\cite{wangExtracellularMatrixSheet2022, liToughCompositeHydrogels2018, nihDualfunctionInjectableAngiogenic2018} and pharmaceutical agents\cite{renAlignedPorousElectrospun2018}, electrical\cite{cheng3DStructuredSelfpowered2020, bergerContinualElectricField1994, wuElectroactiveBiodegradablePolyurethane2016}
, mechanical\cite{wahlstenMechanicalStimulationInduces2021, theocharidisStrainprogrammedPatchHealing2022, guMicroactuatorArrayBased2024}, and electromechanical\cite{taiEnhancedPeripheralNerve2023, sunBioabsorbableMechanoelectricFiber2024, guptaDevelopmentElectroactiveHydrogel2021} stimulation have been established as efficacious therapeutic modalities. These methods expedite the healing process in skin\cite{sunBioabsorbableMechanoelectricFiber2024, blaisSensoryNeuronsAccelerate2014}, nervous\cite{kasubaMechanicalStimulationElectrophysiological2024} , cardiac\cite{stoppelElectricalMechanicalStimulation2016, choiFullyImplantableBioresorbable2021}, and muscular\cite{basurtoAlignedElectricallyConductive2021, guMicroactuatorArrayBased2024} tissues by guiding cellular processes such as migration, proliferation, differentiation, and apoptosis \cite{caoPhysicalCuesScaffolds2024,nardiniSystemicWoundHealing2016}. The effective conveyance of electrical and mechanical stimuli to cultured cells requires direct contact with the target tissue or stimulus transduction facilitated through an engineered construct. For the latter, electroactive polymers (EAPs), which are capable of delivering electromechanical stimulation or interpreting cellular signals, are being increasingly explored for their applications in skin, nerve, muscle, and cardiac regeneration, among others \cite{stewartElectricalStimulationUsing2015,mirandaalarcónUseCollagenMethacrylate2023,wangElectricalStimulationEnhances2023, zareiFabricationCharacterizationConductive2021}. However, conventional electroactive materials exhibit poor biocompatibility, limited degradability, and often require operating conditions that fall outside of the physiological range \cite{wuElectroactiveBiodegradablePolyurethane2016,yoshidaMacroscaleCollagenActomyosinHybrid2023,kimAdvancesBiodegradableSoft2022,kangBionicArtificialSkin2024}. These limitations constrain their utility in regenerative constructs, despite their potential for dynamic control. Some biomaterials exhibit intrinsic electroactivity (such as collagen and hydroxyapatite) through piezoelectricity; nonetheless, both the voltage output and the magnitude of actuation are insufficient for the development of controllable dynamic platforms \cite{kwonPiezoelectricHeterogeneityCollagen2020, burgessTappingTissueBioelectromechanics2024}.

To address this challenge, we present an ionic electrochemical actuator comprised exclusively of biologically-derived constituents, enabling ECM materials to be capable of electromechanical signal translation as a stimulation platform for dynamic tissue engineering. The conjugation of sulfonated biomolecule 2-aminoethanesulfonic acid (commonly called `taurine') to collagen is achieved by employing common cross-linking agents. This side group modification of proteins with a low pKa functional group is hypothesised to facilitate the formation of an ionomers from the base ECM biopolymers. In this work, we show that the development of this net negative charged collagen ionic polymer results in an electromechanically active substrate capable of displacements comparable to those of commercial synthetic EAPs, and surpassing existing biological EAPs. The compatibility, degradation, and sensitivity of the material platform highlight the potential for ECM-derived actuators as a candidate for dynamic tissue engineering constructs. They offer advantages over the currently available ionic polymer actuators for controlled electromechanical stimulation of cells and tissues, promoting regeneration and repair. Our research establishes a novel framework by integrating the electromechanical capabilities of existing EAPs in an ECM-based platform to produce tissue-mimicking matrices. This innovation not only enhances the electromechanical performance but also precisely replicates the biochemical tissue microenvironment, facilitating advanced tissue engineering applications. 

\section{Concept overview and ionomeric extracellular matrix design}

\begin{figure}[ht]
    \centering
    \includegraphics[width=.9\linewidth]{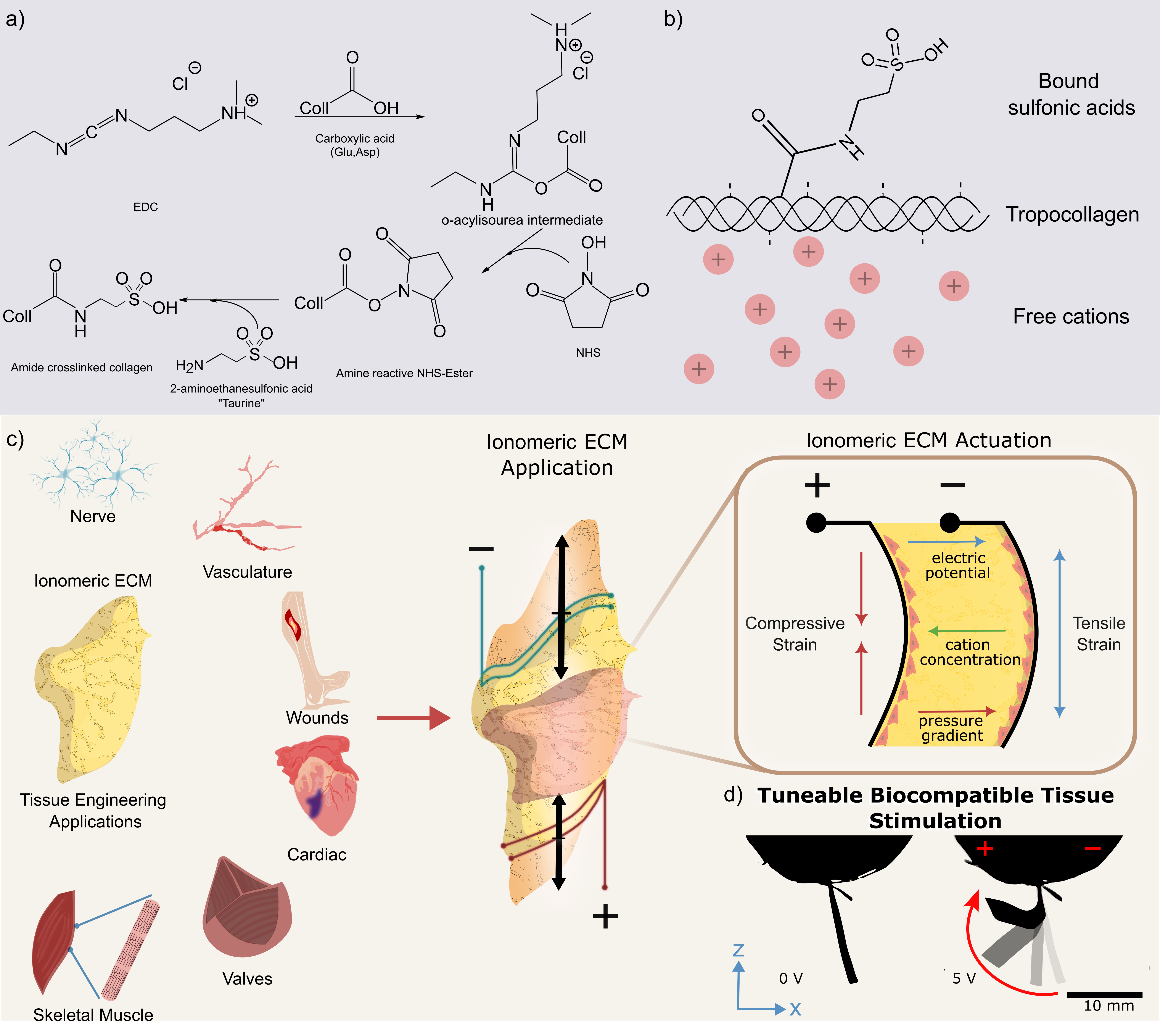}
    \caption{a) Proposed EDC-NHS conjugation mechanism for functionalisation of collagen with a taurine. b) Distribution of the ionomeric extracellular matrix (iECM) consisting of a bound taurine anion linked to the tropocollagen helix in the presence of hydrated `free' cations. c) Dynamic tissue engineering applications for ionomeric ECMs and the proposed mechanism of ionic actuation d) iECM resting position at 0 V and actuation path under 5 V DC over 180 seconds.}
    \label{fig:Overviewconceptandmaterialdesign}
\end{figure}

Therapies based on electrical and mechanical stimulation have demonstrated considerable potential in the realm of tissue engineering, notably in peripheral nerve regeneration, angiogenesis, wound healing, cardiac tissue repair and skeletal muscle regeneration \cite{sunBioabsorbableMechanoelectricFiber2024, blaisSensoryNeuronsAccelerate2014, wuAcceleratedIntestinalWound2024, shouMechanoActivatedCellTherapy2023}. These applications often aim to address complications that lead to inflammatory stagnation in complex and chronic conditions, expediting recovery in nerve conduits\cite{taiEnhancedPeripheralNerve2023, zhangNerveGuideConduits2024}, and improved maturity in stem cell laden scaffolds prior to implantation\cite{shi3DMagneticallyActuated2021, damarajuThreedimensionalPiezoelectricFibrous2017, wangElectrospunConductiveNanofibrous2017}. However, the implementation of these therapies in the development of tissue engineering devices is constrained by the base materials, which frequently comprise synthetic, non-biodegradable, and cytotoxic EAPs. There exists a need for a resorbable and biocompatible EAP to effectively bridge the gap between the functional capabilities of synthetic ionic EAPs such as Nafion\texttrademark~and Flemion\texttrademark, while simultaneously possessing suitable biochemical characteristics that facilitate cellular integration and enable the resorption of the device through direct degradation and remodelling. To tackle these challenges, we have designed an entirely resorbable ionic-EAP, capable of electromechanical actuation, solely derived from ECM-based materials: an Ionomeric Extracellular Matrix (iECM). 

Our iECM is capable of inducing mechanical strain in response to an electrical stimuli due to the conjugation of the sulfonated biomolecule taurine to the tropocollagen helix. It is theorised that the free amine groups of taurine can form covalent amide bonds with the free carboxyl groups, typically found within the GXOGER sequence of collagen \cite{nairCrosslinkingCollagenConstructs2020}, facilitated by chemical linkers such as EDC-NHS. This hypothesis is demonstrated in Fig.\ref{fig:Overviewconceptandmaterialdesign} a--b), which depicts taurine conjugation to the collagen backbone. When subject to an electric potential, the bound anions are attracted to the anode while the free cations in the hydrated substrate migrate towards the cathode, inducing an osmotic potential in the polymer membrane. The electro-osmotic pressure gradient in the iECM drives actuation in the membrane, inducing a strain on the surrounding cells and tissue through electrical and mechanical stimulation in a tissue-mimicking material (Fig.\ref{fig:Overviewconceptandmaterialdesign}c)).

To assess the mechanical response under an applied field, we used a setup that applied a potential across the thickness of the film and two stereo cameras at right angles to record the 3D displacement over a range of voltages and times (Supplementary Figure 1). This experimental configuration enables the assessment of the bending displacement of a freely suspended film, which is constrained by an electrode clip to induce an electric field on both sides. Consequently, the film exhibits bending towards the cathode as a reaction to the electro-osmotic pressure, analogous to synthetic ionic polymer membranes. When normalised by their film lengths, the iECM films fabricated in this work were found to produce a large magnitude of electromechanical displacement under a 5 V DC (Fig.\ref{fig:Overviewconceptandmaterialdesign}d)) than  chitosan- and cellulose-derived films reported in the literature, and a normalised maximum displacement comparable to commercial synthetic EAP-based actuators such as Nafion\texttrademark~ under similar test conditions\cite{panditaManufacturingActuationCharacterization2006, olverabernalChitosanPVANanofibers2023, huangFastresponseElectroactiveActuator2021}. A full comparison of the maximal deflections observed with our iECMs compared to synthetic EAPs and biomaterial-based EAPs is provided in Tab. \ref{tab:EAPcomparison}.

\begin{table}[h]
    \centering
    \caption{Comparison of ionic EAPs actuation performance for a range of synthetic and biopolymeric examples.}
    \label{tab:EAPcomparison}
    \begin{tabular}{>{\centering\arraybackslash}p{3cm}>{\centering\arraybackslash}p{1cm}>{\centering\arraybackslash}p{2.5cm}>{\centering\arraybackslash}p{2.5cm}>{\centering\arraybackslash}p{2.5cm}>{\centering\arraybackslash}p{2.5cm}>{\centering\arraybackslash}p{2cm}}
    \hline
         Ionic EAP&  Electrode Material&  Free Sample Size (L x W x t) (mm)&  Stimulation Parameters&  Maximum Displacement (mm)&  Normalised Maximum Displacement (\% of film length) & Ref.\\
         \hline
 Nafion-117
& Ag& 20 x 5 x0.18& 2 V, DC, 6 sec& -& 90&\cite{panditaManufacturingActuationCharacterization2006}\\
\hline
 Nafion-117& Pd& 30 x 5 x 0.05--0.2& 2 V, DC, 50 sec& 17.4& 58&\cite{wangEffectDehydrationMechanical2014}\\
         \hline
         Aquivion&  Pt
&  48 x 12
&  3 V, AC, 100 mHZ
&  11.02
&  23
& \cite{trabiaSearchingNewIonomer2017}\\
         \hline
         PSMI-incorporated PVDF
&  Pt
&  20 x 5
&  2 V, DC, 30 sec
&  5
&  25
& \cite{luBiomimeticActuatorBased2008} \\
         \hline
         Chitosan/PVA
&  Pt 
&  20 x 4 x 0.03
&  10 V, DC, 50 sec
&  4.2
&  21
& \cite{olverabernalChitosanPVANanofibers2023} \\
         \hline
         Chitosan based film
&  Pt 
&  40 x 10
&  7 V
&  17
&  42.5
& \cite{cuellar-monterrubioDevelopmentElectroactiveBiopolymerbased2022} \\
         \hline
         TEMPO oxidized nanocellulose
&  Au
&  40 x 10
&  10 V, 100 sec
&  32.1
&  80
& \cite{huangFastresponseElectroactiveActuator2021} \\
         \hline
         Chitosan-MWCNT-GO
&  GO
&  39.76 x 7.88 x 0.023
&  
3 V, 250 sec&  5.65
&  14
& \cite{jiaStudyActuationCharacteristics2022} \\
         \hline
         Collagen-Taurine iECM
&  Ag
&  15 x 5
&  5 V, DC, 180 sec
&  15
&  100
& This work \\
         \hline
    \end{tabular}
\end{table}

The promising electromechanical properties of iECM films highlight their potential utility in applications beyond standard synthetic EAP-based actuators, particularly where biocompatibility is crucial across a variety of tissue engineering platforms. A key characteristic observed across human dermal fibroblasts (HDFs), mouse myoblasts (C2C12), and human umbilical vein endothelial cells (HUVECs) is the comparable or elevated compatibility exhibited for our collagen-taurine-EDC/NHS linked (CTE) iECM films, as compared with uncrosslinked collagen (C), cross-linked collagen (CE), and unlinked collagen-taurine (CT) films over a seven-day culture duration.

Thus, by leveraging conventional carbodiimide linkers to bind charged biomolecules to a biopolymer, we have manufactured an iECM capable of producing macroscopic deflections under applied electromechanical stimulation in a tissue-engineered construct. 

\section{Enhancing the electroactive performance of iECM films}

\begin{figure}[ht]
    \centering
    \includegraphics[width=.87\linewidth]{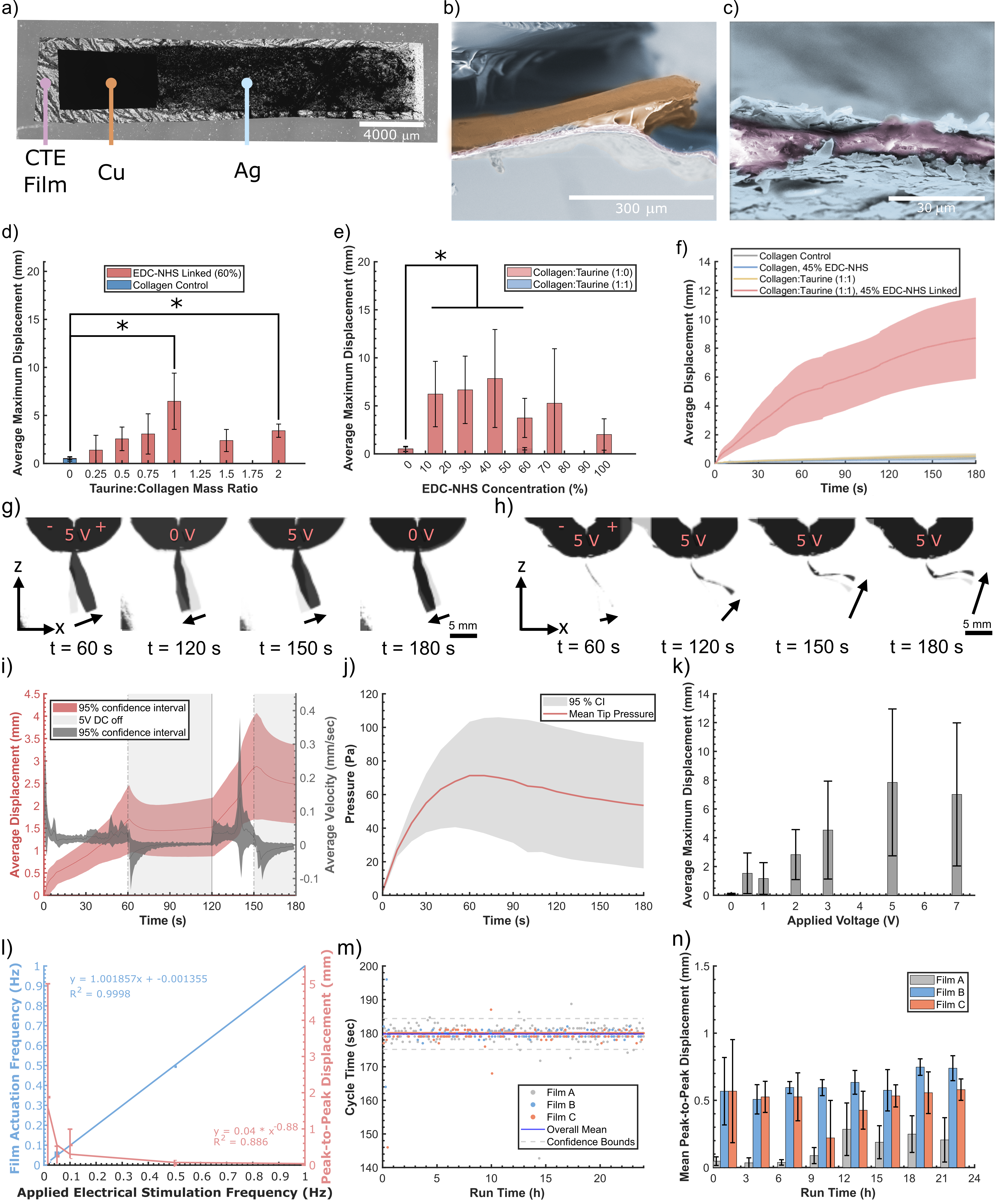}
    \caption{a) Electroactive CTE films coated with silver electrodes and copper contact pads. b) Cross-section of a copper contact tab interface at the CTE film surface. c) Cross-section view of silver plated electrodes on either side of a CTE film. d) Tip displacements as a function of the taurine concentration variation and e) EDC-NHS linking solution concentrations. f) Comparison of the average displacement-time curves for the most optimum formulation of 0.5 wt.\% collagen with a taurine-collagen mass ratio of 1 and an EDC-NHS solution concentration of 45\%. g) Characteristic film actuation and relaxation response following electrical stimulation at 5 V for 60 seconds, followed by a 60 second rest period before a second stimulation and rest cycle of 30 seconds. h) Characteristic film actuation response at 5 V for 180 seconds. i) Characteristic displacement and velocity response of optimised CTE films to an applied 5 V DC electric potential. j) Output pressure generated from optimised CTE films over 180 seconds at a 5 V potential. k) Average maximum film displacements after 180 seconds from 0--7 V sweep. l) Actuation frequency of thin CTE films when exposed to a positive 5 V square wave pulse from 15 mHz to 1 Hz. m-n) Cycle time and peak to peak displacement magnitude of CTE films from a triplicate actuation stability study where a voltage cycle of 180 seconds (60-second activation at 5 V and 120-second activation at 0 V was applied over 24 hours. All individual cycle times are plotted, highlighting the mean of 179.8 $\pm$ 2.3. Statistical significance ($p < 0.05$) against the control samples is indicated by a * following a Kruskal-Wallis and a post hoc Dunn’s test.}
    \label{fig:ElectroactiveOptimisationAndPerformance}
\end{figure}

The composition of iECMs were investigated as a function of varying linker and sulfonic acid concentrations to identify an enhanced region of electrochemical actuation. 5-by-20 mm strips of collagen films were assembled in a tri-layered configuration (Fig.\ref{fig:ElectroactiveOptimisationAndPerformance}.a--c)), comprising of silver conductive sheets with a thickness of 2.18 $\pm$ 1.00 $\mu$m, plated on either side of a collagen film with a thickness of 10.03 $\pm$ 3.38 $\mu$m (Supplementary Fig.2). Copper contact pads were incorporated to ensure consistent loading position between samples. A DC electric potential of 5 V was applied through the silver electrodes, generating an electric field along the length and across the thickness of the film (Supplementary Fig.3). 

iECMs were designed with the goal of creating a material that supports dynamic tissue engineering. Stimulation-based therapies aiming to promote healing are based on the cell, tissue, and organ response to a force (strain inducing a pressure on the surrounding tissue) and current or voltage. These therapies work to activate/deactivate genes and proteins to promote cell proliferation, migration, nutrient diffusion, angiogenic, and other desirable pathways associated with accelerated tissue repair \cite{caoPhysicalCuesScaffolds2024, chenElectricalStimulationNovel2019}. Such stimuli often operate within an optimal range; therefore, the tunability of the mechanical displacement and subsequent force output to electrical stimuli are important factors to consider when a platform incorporating multiple exogenously deliverable stimuli is developed. Therefore, electromechanical actuation test conditions were maintained at physiologically relevant relative humidity ($>$ 80\%) or under hydrated conditions and at room temperature to mimic potential in-vitro and in-vivo applications. 

The tip displacement was assessed over time for films with 15 mm of free length, following hydration in 1$\times$ PBS to simulate physiological ion concentrations. An initial increase in taurine content to a 1:1 collagen:taurine mass ratio enhanced the average maximum actuation displacement to a maximal deflection of 6.5 mm, a 13-fold statistically significant increase compared to collagen controls, whereas further taurine increases beyond this ratio diminished the actuation magnitude. The linker solution was a 3 : 1.8 : 1 EDC: NHS : Collagen molar ratio, henceforth referred to as the `60\%' EDC-NHS condition (Fig.\ref{fig:ElectroactiveOptimisationAndPerformance}.d)). The effect of EDC-NHS linker concentration for a 1:1 collagen:taurine mass ratio is shown in Fig.\ref{fig:ElectroactiveOptimisationAndPerformance}.e) . It can be seen that further electromechanical actuation enhancement was achieved when we altered the EDC-NHS linker concentration, from 15--100\%. A broad region of statistically significant electromechanical enhancement was present from 15--60\% EDC-NHS, with a peak average displacement of 7.8 mm at 45\% EDC-NHS, although no statistically significant difference was observed between taurine-loaded and EDC-NHS linked films (Fig.\ref{fig:ElectroactiveOptimisationAndPerformance}.e)). Hence, it is conceivable that a dual electroactive and mechanical enhancement could be exploited by altering the crosslinking concentration within this actuation region for EDC-NHS. By varying sulfonic acids (taurine) and linker (EDC-NHS) compositions, we identified an optimal actuation region for a 0.5 wt.\% collagen film with a 1:1 taurine mass ratio and 45\% EDC-NHS concentration, resulting in a 19.4 times greater enhancement in deflection over collagen controls (C), unlinked collagen-taurine (CT), and cross-linked collagen (CE) (Fig.\ref{fig:ElectroactiveOptimisationAndPerformance}.f)).

Displacement and force are vital for tissue engineering constructs, as they induce strains and pressures that activate mechanoreceptors in cell membranes, affecting cell behaviour in growth, apoptosis, and differentiation pathways \cite{burgessTappingTissueBioelectromechanics2024}. Thus, controlled response rate and pressure generation are essential material characteristics. The material demonstrates the capacity for both unidirectional and multidirectional motion, as shown in Fig.\ref{fig:ElectroactiveOptimisationAndPerformance}.g--h. Under a 5 V electric potential, displacement is directed towards the anode during activation periods of 60 seconds and 30 seconds. Upon removal of these stimuli, relaxation towards the cathode is observed for most samples (Fig.\ref{fig:ElectroactiveOptimisationAndPerformance}.i.).  As the driving force for ion migration is removed, a counter-motion is expected as the system attempts to reach an ion equilibrium in the membrane. For samples that maintained their final position with minimal relaxation, a combination of film drying accompanied by slow ion migration in the absence of an electric field is expected to contribute to the observed response. In the present configuration, the application of a 5 V DC potential enables the CTE actuators to produce an average peak actuation pressure of 71 Pa at the film's tip (Fig.\ref{fig:ElectroactiveOptimisationAndPerformance}.j.). 

Next, we measured the enhanced collagen-based iECM response to a range of activation voltages (0--7 V). The magnitude of displacement scaled with the applied voltage from 1.5 $\pm$ 1.4 mm at 0.5 V to a peak of 7.9 mm at 5 V (Fig.\ref{fig:ElectroactiveOptimisationAndPerformance}.k).The displacement magnitude observed at 5 V and 7 V exhibited no significant variation. During the displacement assessment at 7 V, an accelerated drying of the film was noted, which may contribute to the restricted actuation displacement observed with the increased voltage. 

We further explored the sensitivity to the actuation stimuli by applying square wave voltages from 15 mHz to 1 Hz. A linear response between the applied stimulation frequency of a 5 V square wave electric potential and the actuation frequency of collagen-based iECMs can be observed from the experimental data shown in Fig.\ref{fig:ElectroactiveOptimisationAndPerformance}(k) (R$^{2}$ = 0.9998) with negligible variability across samples in the tested range (15 mHz to 1 Hz). Assessment of peak-to-peak displacement across varying stimulation frequencies indicated a decrease as the applied frequency increased. This decrement is not characterised by linearity but rather approximates a power law relation (R$^{2}$ = 0.886). The non-linear response observed in ionic actuators, as presented by \citeauthor{kotalHighlyBendableIonic2018}, is proposed to result from a combination of electrode asymmetry and the dynamics of ion diffusion and charge accumulation \cite{kotalHighlyBendableIonic2018}. An additional sweep was applied to an electrode plated CTE film to assess the impact of a shift in stimulation frequency from 5 Hz to 25 mHz on the actuation oscillation frequency. As shown in Supplementary Fig.4, the material remains sensitive to the AC signal over this frequency range and the peak-to-peak displacement magnitude decreases as frequency increases. Therefore, iECMs exhibit a reliable and repeatable frequency response to electrical stimuli. Over a 24 hour AC-stimulation experiment, the films demonstrated sustained response to repeated activation cycles, consisting of a 120-second 5 V square wave potential followed by a 60-second rest phase. The films were sensitive to the applied stimulus during the test period, maintaining an average peak-to-peak displacement cycle time of 179.8 $\pm$ 2.3 seconds and a sustained peak-to-peak displacement magnitude (Fig.\ref{fig:ElectroactiveOptimisationAndPerformance}g--h)) over the duration of the 24 hour experiment, representative actuation cycles are shown in supplementary Fig.5.

\section{Characterisation of collagen based iECMs}

\begin{figure}[ht]
    \centering
    \includegraphics[width=0.87\linewidth]{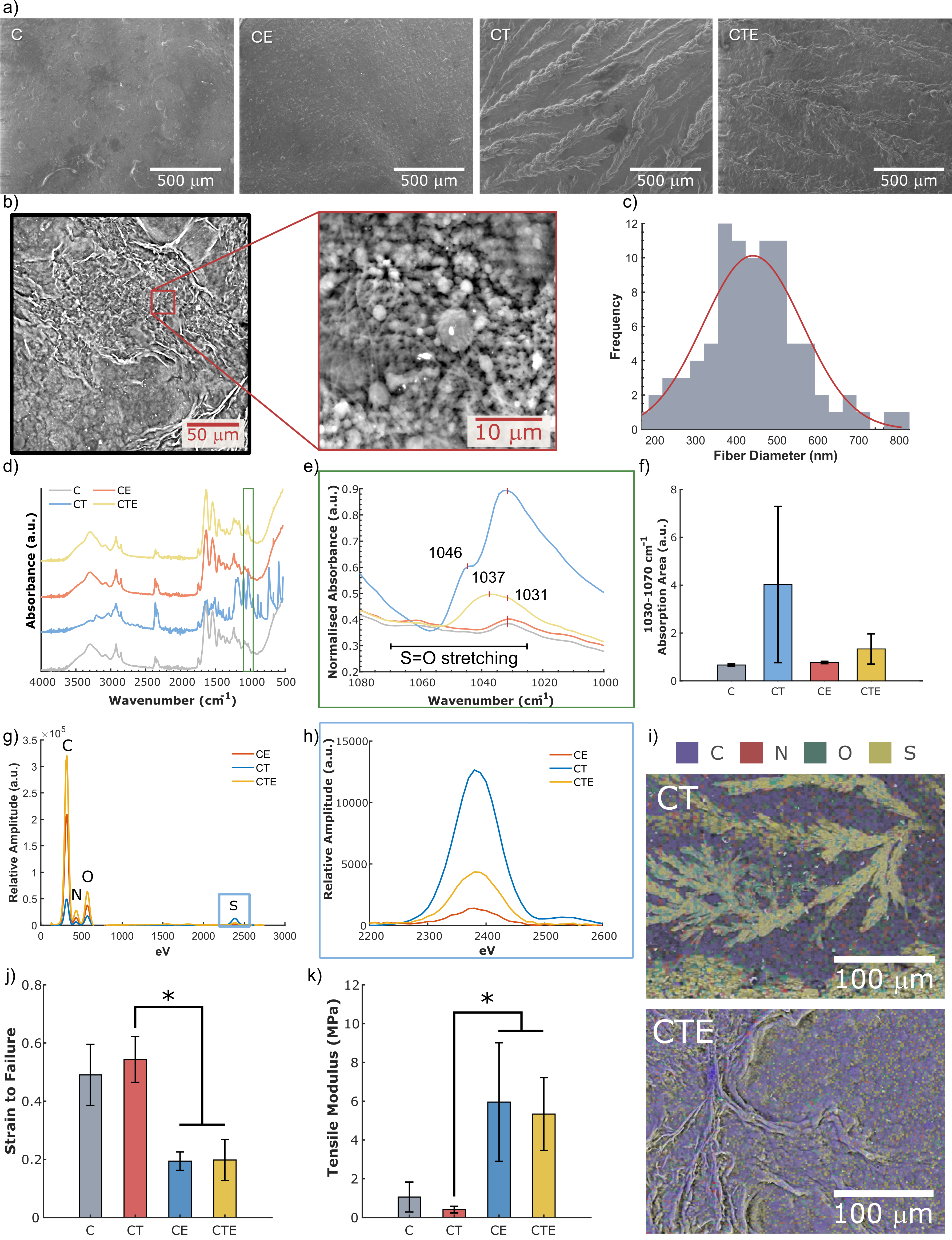}
    \caption{Physical and chemical characterisation of electromechanically enhanced collagen films. a) SEM micrographs illustrating the surface morphology of the films. b) SEM micrographs that emphasise the fibre topography within CTE films. c) Distribution of fibre diameters in the CTE matrix. d) FTIR spectra ranging from 500 to 4000 cm$^{-1}$. e) Enlarged spectral region from 1030 to 1070 cm$^{-1}$, highlighting S=O stretching and alterations in peak positions subsequent to collagen functionalisation with taurine. f) Mean absorption area between 1030 to 1070 cm$^{-1}$ across triplicate films. g) EDS spectra identifying the peaks corresponding to carbon, nitrogen, oxygen, and sulfur. h) Enlarged sulfur region from the EDS spectra demonstrating increased sulfur peak intensity in the CT and CTE films compared to the CE control. i) EDS mapping delineating sulfur-rich regions in yellow. j--k) Evaluation of strain to failure and tensile modulus for C, CT, CE, and CTE films. Statistical significance ($p < 0.05$) against the control samples is indicated by a * following a Kruskal-Wallis and a post hoc Dunn’s test.}
    \label{fig:EnhancedMaterialProperties}
\end{figure}

The cellular response to tissue engineering constructs is strongly impacted by the physicochemical properties of the substrate. As a result, we investigated the hydrophobicity, morphology, topography, macroscopic mechanical response, and chemical composition of the iECM films. 

Changes in surface morphology following the immersion-linking process were apparent in SEM images (Fig.\ref{fig:EnhancedMaterialProperties}.a) of collagen films. Although collagen and cross-linked collagen retained similar surface characteristics, the addition of taurine induces the formation of dendrites due to phase separation during the drying phase following drop casting. When subjected to EDC-NHS-mediated linking, the prevalence of dendritic spines is reduced, indicating the loss of any unbound taurine. 

 In CTE iECMs, fibre bundles of 440 $\pm$ 120 nm can be observed as shown in the SEM micrographs in Fig.\ref{fig:EnhancedMaterialProperties}.b--c). This formation of bundled fibres is characteristic of EDC-NHS crosslinking in collagen, and is within the range of fibre bundles found in crosslinked collagen films \cite{nairSelfassemblyCollagenBundles2019}. Since these are uncommon in films\cite{nairSelfassemblyCollagenBundles2019}, taurine conjugation can be interpreted as competing with EDC-NHS-mediated collagen cross-linking during the electroactive functionalisation process.

Moisture retention is vital for electromechanical actuation and cell survival. Collagen control films swelled by 67.0\% $\pm$ ± 6.9\%. Taurine addition and 45\% EDC-NHS crosslinking reduced swelling to 49.0\% $\pm$ ± 3.3\% and 55.2\% $\pm$ ± 6.4\%, respectively. Collagen-taurine samples linked with 45\% EDC-NHS swelled 67.8\% $\pm$ ± 5.8\%, a change of less than 1\% from the control (Supplementary Fig.6). The net mass of the CTE films was found to be higher than that of C and CE controls but less than that of unwashed taurine-collagen CT films. This supports the retention of taurine after conjugation, with unlinked taurine dissolving in the medium.

By choosing collagen, a protein with less than 1\% endogenous sulfur in its amino acid sequences\cite{eastoeAminoAcidComposition1955}, we are able to delineate the inclusion of free and conjugated taurine through a range of vibrational and elemental spectroscopic techniques. We confirmed the change in the chemical composition of the collagen-taurine films through Fourier transform infrared (FT-IR) spectroscopy (Fig.\ref{fig:EnhancedMaterialProperties}.d) and energy-dispersive X-ray spectroscopy (EDS) to identify the change in CTE composition as compared with controls: C, CT and CE. 

Peaks corresponding to sulfonic groups are observed in the conjugated material. The most prominent change, after taurine conjugation, was observed in the 1030--1070 cm$^{-1}$ band, with CTE peaks shifting towards 1037 cm$^{-1}$ corresponding to S=O bond stretching\cite{maCharacterizationSeparationPurification2018} (Fig.\ref{fig:EnhancedMaterialProperties}.e)). The area under the curve in this region was calculated to determine the relative sulfur content across the functionalised material (CTE) and the respective controls  following a baseline normalisation according to the local minima of the curve. We observed a 600\% increase in peak area for CT and 200\% increase in peak area for CTE as compared with CE and C controls, confirming that taurine is retained within the matrix due to the hypothesised conjugation (Fig.\ref{fig:EnhancedMaterialProperties}.f). In all films, the characteristic amide I (1631 cm$^{-1}$), II (1550 cm$^{-1}$) and III (1270 cm$^{-1}$) peaks of collagen (Fig.\ref{fig:EnhancedMaterialProperties}.d) are unchanged suggesting no alteration of the underlying triple helical structures\cite{terziEffectsProcessingStructural2018}.

Elemental analysis corroborates the EDC-NHS mediated linkage of sulfonic acids to the collagen matrix, revealing an increase in sulfur content (emission peaks at 2390 eV in Fig.\ref{fig:EnhancedMaterialProperties}g--h) in the functionalised material. The peak amplitude of the functionalised material corresponds to a 2.3-fold increase in sulfur content following taurine conjugation, retaining 8.8 \% of the initial loading concentration (see Tab. \ref{tab:EDSAtomicPercentages}). The spatial distribution of sulfur (yellow pixels in Fig.\ref{fig:EnhancedMaterialProperties}.i), in the CT films highlights a taurine-rich dendritic network with relatively low sulfur concentrations in the biopolymer matrix prior to linking. Following the conjugation process, the sulfur concentration in the dendrites appears to be reduced, and is found to be distributed more evenly across the surface of the film, barring a few pockets of high sulfur concentrations at the dendrite-polymer interface. We confirm this shift in spatial distribution by assessing the atomic percentage of sulfur at the independent regions using laser induced breakdown spectroscopy (LIBS); see Supplementary Fig.7. 

\begin{table}[h]
    \centering
    \caption{Atomic concentration percentage
quantification from EDS of thin films.}
    \begin{tabular}{|>{\centering\arraybackslash}p{5.25cm}|>{\centering\arraybackslash}p{5.25cm}|}
    \hline
         Sample&  S (at.\%)\\
         \hline
         CE&  0.13
\\
\hline
         CT&  3.40
\\
\hline
         CTE&  0.30
\\
\hline
    \end{tabular}
    \label{tab:EDSAtomicPercentages}
\end{table}

The incorporation of taurine anions into the collagen matrix was confirmed by measuring the charge density of cast films using the Teorell-Meyer-Siever (TMS) theory \cite{pichenyMethodsMeasurementCharge1976}. Samples were placed between two compartments with equal volumes of 1 X PBS, and sodium chloride solutions of varying concentrations were added. The resulting potential difference between compartments, indicative of the sample's immobile charge density, was analysed using TMS theory, as shown in Eq.\ref{eq: Charge density voltage terms} and presented in Tab.\ref{tab:charge_density}. Full calculations are provided in the supplementary material, section I. Tab. \ref{tab:charge_density}
demonstrates the nearly tenfold increase in the magnitude of charge density following functionalisation with sulfonic acids. In contrast, the CT films failed to achieve a steady-state potential due to the leaching of unbound taurine.
\begin{equation}
\label{eq: Charge density voltage terms}
    V_0=\Delta\Phi_\text{diff}+(\Delta\Phi^{'}_{D}-\Delta\Phi^{''}_{D})-\frac{RT}{F}\ln(\frac{a^{''}_{Cl}}{a^{'}_{Cl}})
\end{equation}

\begin{table}[htbp]
    \centering
    \caption{Charge density results form C, CT and CTE films.}
    \label{tab:charge_density}
    \begin{tabular}{|>{\raggedright\arraybackslash}p{2cm}|p{2.5cm}|p{2cm}|p{2cm}|p{2cm}|}
        \toprule
        Sample & Electrolyte  Concentration Gradient & Electrolyte Volume (ml) & Measured Voltage (mV) & Magnitude of Charge Density $|\rho_m|$ (C/m$^3$)\\
      \hline
        C   & 2 & 100 & 16.94 & $1.31\times 10^6$ \\
        \hline
        CT  & 10 & 100 & --- & --- \\
        \hline
        CTE & 10 & 100 & 5.55 & $11.69 \times 10^6$\\
        \hline
    \end{tabular}
\end{table}

The ultimate tensile strength (UTS) of collagen films is reported as 2.30 $\pm$ 1.99 MPa, with the tensile modulus (from the initial linear region of the stress-strain curve, see supplementary Fig.8) of 1.06 $\pm$ 0.77 MPa and a strain to failure of 49 $\pm$ 11\%. CT films exhibited no significant differences across all three parameters, with values of 1.60 $\pm$ 0.48 MPa, 0.4 $\pm$ 0.17 MPa, and 54 $\pm$ 8\% for UTS, tensile modulus, and strain to failure, respectively (Fig.\ref{fig:EnhancedMaterialProperties}.j--k). Films cross-linked with 45\% EDC-NHS displayed an increased UTS of 3.03 $\pm$ 0.80 MPa and demonstrated statistically significant changes in strain to failure (19 $\pm$ 3\%) and tensile modulus (5.96 $\pm$ 3.06 MPa) compared to CT at the 5\% significance level, and compared to both CT and C films at the 10\% level (for strain to failure). CTE films do not significantly differ from CE films, but are statistically significant against CT and C ($p < 0.05$) with UTS recorded at 3.09 $\pm$ 1.84 MPa, tensile modulus at 5.33 $\pm$ 1.88 MPa, and a strain to failure of 20 $\pm$ 7\%. The mechanical properties of CTE films are, therefore, deemed suitable for applications in tissue engineering, given that their properties have not been significantly altered from crosslinked collagen, an ubiquitous material in tissue-engineered constructs.

\section{iECM \textit{in vitro} Biocompatibility }

\begin{figure}[ht]
    \centering
    \includegraphics[width=1\linewidth]{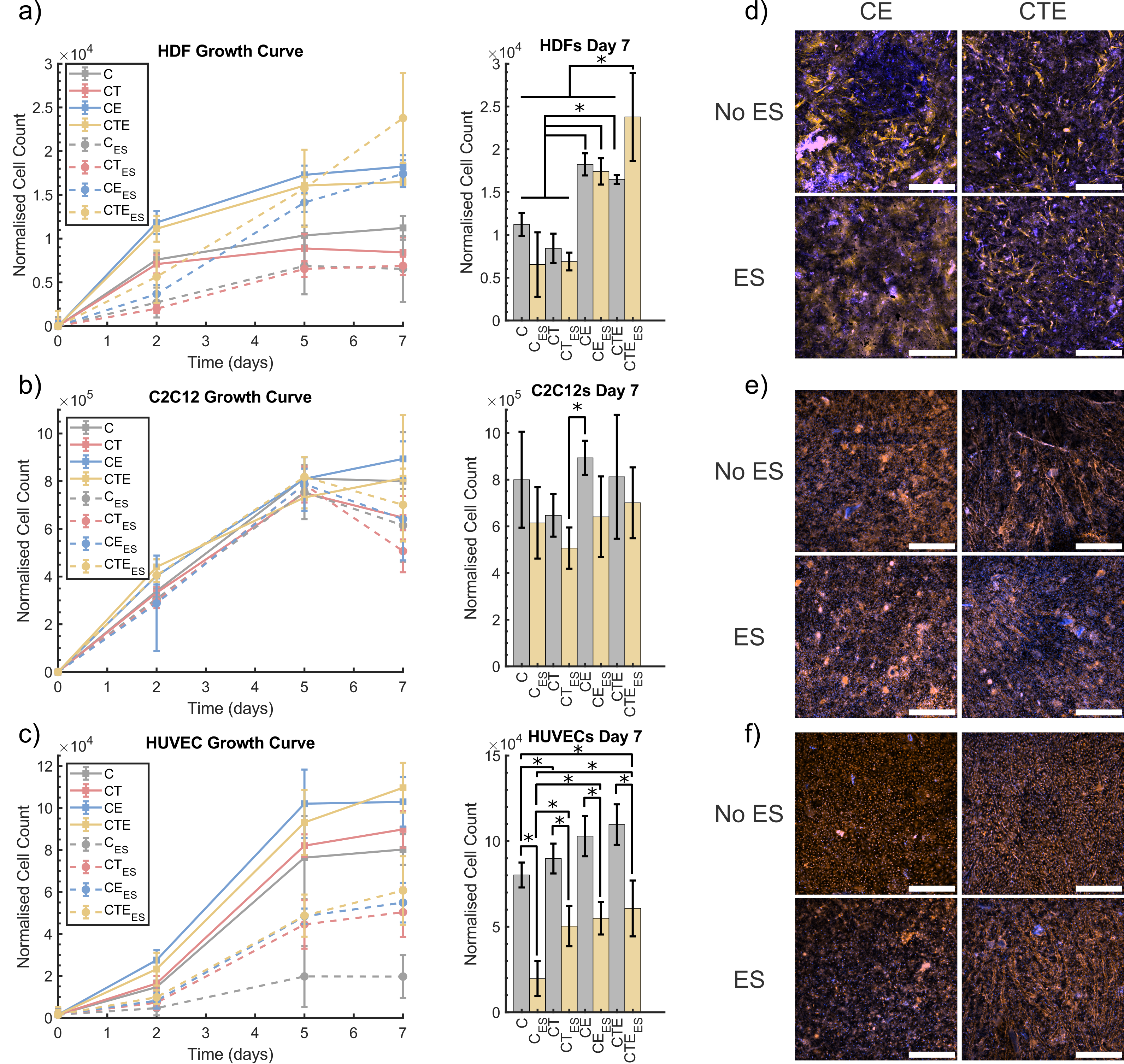}
\caption{The HDF (a), C2C12 (b) and HUVEC (c) cell count in both stimulated and non-stimulated films, including C, CT, CE, and CTE over 7 days. d) Fluorescent staining for cell nuclei (DAPI) and cytoskeleton (Phalloidin) was conducted to illustrate cell confluence and distribution on day 7 across both stimulated and non-stimulated CE and CTE groups in HDFs (d), C2C12s (e) and HUVECs (f), scale bar = 500 $\mu$m. Statistical significance ($p < 0.05$) is indicated by a * following a one-way analysis of variance test.}
    \label{fig:BiocompatibilityAndRegnerativePotential}
\end{figure}
Electrical stimulation impacts cell behaviour through multiple pathways, and has been used concurrently to both enhance cell proliferation for applications such as wound healing and angiogenesis, but also reduce cell proliferation in applications such as keloid scarring. The choice of stimulation parameter depends on both the exact cell types, substrates, and electrode geometries, and can be tuned for the application of interest. Here we demonstrate an exemplar voltage of 450 mV/mm, to understand the additional impact on cell viability and proliferation provided by an electroactive substrate when stimulated electrically in C, CT and CE films and electromechanically in our CTE iECM across three distinct cell lines: HDFs, C2C12s, and HUVECs. Cell lines were cultured on C, CT, CE and CTE films over 7 days with and without electrical stimulation with Presto Blue assays conducted on days 0, 2, 5 and 7 (Fig.\ref{fig:BiocompatibilityAndRegnerativePotential}). 

\subsection{Platform material compatibility}

We established that the linkage of taurine to the collagen matrix had no significant effect on cellular biocompatibility among the three cell lines, indicating that the attachment of taurine to the collagen membrane and the alteration of the charge density of the newly electroactive biopolymer do not affect the cell population over the duration of a 7-day study. Furthermore, both HDFs and HUVECs exhibited a marked increase in proliferation when cultured on CE and CTE films that produced cell counts 1.6 and 1.5 times higher than collagen controls for HDFs and 1.3 and 1.4 times higher for HUVECs, respectively. The incorporation of charged groups via bound taurine did not markedly alter the cellular response to the CE films in the absence of electrical stimulation. Throughout the 7-day study period, all non-stimulated cultures exhibited an initial rise in cell count, followed by a plateau between days 5 and 7.

\subsubsection{Impact of Stimulation on Cell Viability}
We further assess the effects of electrical stimulation on non-functionalised films and electromechanical stimulation on functionalised CTE films at a 4.5 V AC potential with 100 ms pulse widths of a square wave at 1 Hz. This particular stimulus was selected due to the iECM's actuation capabilities observed in electromechanical displacement study, rather than as a parameter for enhancing cell proliferation. Consequently, the application of electrical stimuli in the HDF and HUVEC cultures reduced the initial cell growth curves. However, after the 7-day culture period, no significant differences were recorded between the non-stimulated groups and the electrically stimulated groups for the C, CT and CE films. Introduction of the mechanical stimuli from the CTE ionic actuation did not significantly alter the C2C12 population over the entire culture period but did significantly increase the day 7 cell count for HDFs, suggesting that the added mechanical stimuli enhanced cell proliferation, increasing the cell count by a factor of 1.4 compared to non-stimulated CTE films and 3.6 compared to electrically stimulated HDFs on collagen films. This shows the potential for the iECM films in their current conformation for a population accelerating bioreactor platform. Furthermore, HUVEC viability declined with the additional electrical stimulus. Nonetheless, no significant differences were noted between the CE and CTE films, indicating that the additional mechanical stimuli did not significantly affect the cell populations. The fluorescent images of the CE and CTE films reveal the spatial distribution of cells on both the stimulated and non-stimulated membranes, illustrating the effect on cellular arrangement. The structure and distribution of HDFs exhibit considerable similarity across both types of substrates. However, C2C12s and HUVECs demonstrate a tendency to cluster around the taurine dendrites within the films, with this clustering being more pronounced in samples subjected to electromechanical stimulation. Despite the absence of significant changes in cell confluence for HDFs and C2C12s, a discernible reduction in cell numbers is observed in the HUVEC cultured films. The application of mechanical and electrical stimuli had varying effects on different cell types. While electromechanical stimuli enhanced HDF proliferation significantly in CTE films, no significant change was observed in C2C12 populations. For HUVECs, the additional electrical stimuli led to a decrease in cell viability. Despite the presence of aligned structures in some stimulated samples no overall cell alignment was observed across all cell types (supplementary Fig.9). These findings highlight the potential of iECM films for bioreactor applications, particularly to enhance HDF proliferation. 

Thus, we confirm that the collagen-based iECM is compatible with skin fibroblast, skeletal muscle and vascular cell lines. In the absence of electrical stimulation, the cell population is not significantly different from cross-linked collagen. However, introduction of electromechanical stimuli boosts the number of HDFs, exclusively in the iECM films, after 7 days. These findings indicate that our engineered materials offer the potential to accelerate the process of wound healing by facilitating the growth of HDFs. The enhanced electromechanical activity of collagen-based iECMs also highlights the potential for dynamic tissue engineering constructs. In their current configuration, the actuators can generate a mean maximum actuation pressure of 71 Pa at the tip of the film (Fig.\ref{fig:ElectroactiveOptimisationAndPerformance}(l)) and achieve large displacements of up to 100\% relative to the length of the film under low driving voltages (5 V). Additionally, their consistent response to stimulation frequency and ability to maintain this sensitivity for 24 hours illustrates their suitability for sustained mechanical actuation. This makes iECMs promising candidates for future applications as dual-stimulating devices that aim to enhance tissue regeneration. Furthermore, we illustrate the capability to replicate this enhanced electromechanical activity within lyophilised scaffolds, attaining a marked improvement in actuation performance (Supplementary Fig.10), thereby reinforcing their suitability for incorporation into standard tissue engineering constructs.

In summation, the data presented in this paper introduces the concept of iECMs and their capacity to achieve macroscopic electromechanical strains in ECM-derived substrates through the conjugation of sulfonated biomolecules such as taurine, directly to the protein backbone. In combination with our \textit{in vitro} study, this work underscores the potential applications of these iECMs in tissue engineering, enhancing HDF cell populations by a factor of 1.4 in unstimulated samples and by 3.6 when electromechanically stimulated. Thus, iECMs present an opportunity for dual electrical and mechanical stimulation on a tissue mimicking substrate for the first time, expanding the realm of dynamic functionality for future tissue-engineered devices.

\section{Methods}
\subsection{Fabrication}
Suspensions of type I Achilles bovine collagen (Sigma-Aldrich, C9879) and taurine (Sigma-Aldrich, T0625) were prepared by hydration in 0.05 M acetic acid (Thermo Fisher Scientific, 13552310) on roller bed at 4 \degree C for 12 hours, followed by homogenisation. Hydrated slurries were mixed on ice using a T 10 basic ULTRA-TURRAX homogeniser for 10 minutes. 5 g of homogenised mixed slurries were drop-cast onto 5cm-by-5cm square anti-static weighing boats and left to dry for 48 hours before linking. 

Herein, collagen concentrations are referred as a standard whereby X\% represents X g of collagen in 100 mL of acetic acid. Taurine concentration is referred to as mass ratios with respect to collagen (i.e., a taurine mass ratio of 1.5 is equal to 1.5 $\times$ g for that collagen concentration).

\subsubsection{Collagen-Taurine Conjugation}

\paragraph{EDC-NHS}

A linking solution was prepared by mixing 1-ethyl-3-(-3-dimethylaminopropyl) carbodiimide hydrochloride (EDC, Thermo Fisher Scientific) and N-hydroxysuccinimide (NHS, Thermo Fisher Scientific) and collagen at the specific mass ratio of 1.15 : 0.276 : 1 (EDC:NHS:collagen) in 75\% ethanol. This is considered the `100\% concentration solution' and is derived from the molar ratio (5:2:1) of the reactive groups (EDC:NHS:COO$^-$) which assumes 1.2 moles of collagen contain a mole of carboxylate groups (COO$^-$) \cite{oldedaminkCrosslinkingDermalSheep1996,nairCrosslinkingCollagenConstructs2020}. Films were fully immersed in the linking solution for 2 hours at room temperature under constant mixing \textit{via} a shaker plate at 240 rpm.

Collagen (C), unlinked collagen-taurine (CT), EDC-NHS crosslinked collagen (CE) and EDC-NHS linked collagen-taurine (CTE) films were produced for comparative analysis, with C, CE and CT as controls to validate the performance of CTE.

All samples were washed thrice following immersion linking using DI-H$_2$O (5 minute soak, followed by a quick rinse).

\subsubsection{Application of electrodes}
Films measuring 10 mm by 30 mm were coated on both sides with 100 $\mu$l of silver conductive paint (Electrolube, RS Components) using a porous sponge. These films were subsequently sectioned into strips of 5 mm by 20 mm and the edges were trimmed to prevent direct contact with the electrode, as validated by resistance measurements greater than k$\Omega$ across the film's thickness, and less than k$\Omega$ range along its length.

\subsubsection{Electroactive Enhancement Studies of Drop Cast Films}
Two studies assessed the influence of taurine composition and concentration of the linking solution against electroactive displacement of thin film actuators at 5 V for 3 minutes. Study 1 considered the taurine concentrations at mass ratios, relative to collagen, of 0.00, 0.25, 0.50, 0.75, 1.00, 1.50 and 2.00 with 0.5 wt.\% collagen films and a EDC-NHS linking solution of 60\%. Study 2 optimised the linking solution concentration with EDC-NHS solution concentrations of 0, 15, 30, 45, 60, 75 and 100\% in 0.5 wt.\% collagen films with taurine:collagen mass ratios of 1:1.

The various concentration of collagen, taurine and EDC-NHS used in this study can be seen in Tab. \ref{tab:ConditionsOfStudy}. 
\begin{table}[h]
    \centering
\caption{Concentration of polymer films in each optimisation stage}
\label{tab:ConditionsOfStudy}
    \begin{tabular}{>{\raggedright\arraybackslash}p{1.5cm}>{\centering\arraybackslash}p{1.5cm}>{\centering\arraybackslash}p{1.5cm}>{\centering\arraybackslash}p{1.5cm}} \hline
             Material Code&  Taurine Mass Ratio&  EDC-NHS Concentration (\%)& Acetic Acid Concentration (M)\\ \hline
     \multicolumn{4}{c}{Study 1: Influence of Taurine Content on Actuation}\\ \hline
             C1&  
0&  0& 0.05\\ 
             C1E60&  0
&  60& 0.05\\ 
     C1T025E060& 0.25
& 60&0.05\\ 
     C1T050E060& 0.5
& 60&0.05\\ 
     C1T075E060& 0.75
& 60&0.05\\ 
     C1T100E060& 1
& 60&0.05\\ 
     C1T150E060& 1.5
& 60&0.05\\ 
     C1T200E060& 2
& 60&0.05\\ \hline
     \multicolumn{4}{c}{Study 2:  Influence of Linker Concentration on Actuation}\\ \hline
     C1& 0& 0&0.05\\ 
     C1T100& 1& 0&0.05\\ 
     C1T100E015& 1& 15&0.05\\ 
     C1T100E030& 1& 30&0.05\\ 
     C1T100E045& 1& 45&0.05\\ 
     C1T100E060& 1& 60&0.05\\ 
     C1T100E075& 1& 75&0.05\\ 
     C1T100E100& 1& 100&0.05\\
    \end{tabular}

\end{table}

\subsection{Electroactivity Characterisation}

\subsubsection{DC Electroactive Displacement}
All films were hydrated by immersion in 1$\times$ PBS (Thermofisher Scientific inc.) for 5 minutes. Prior to electroactive testing, hydrated films were blotted with an absorbent paper towel to remove surface water.

Hydrated films were clamped between copper electrodes, with 15 mm free length from the clamped end. A 5 V DC potential was applied at room temperature between two 2.5 mm by 2.5 mm copper electrodes. Tip displacement was tracked over 3 minutes \textit{via} two stereo cameras, fixed at 90\degree~to each other recorded film displacement from the $xz$ and $yz$ perspective, respectfully see supplementary Fig.1.
A custom-built humidity chamber was maintained at $>$80\% relative humidity (RH) for all experiments. 

\subsubsection{\label{DPT}Dynamic Pixel Tracking for Deformation Analysis}
Videos were analysed using a MATLAB script to track a single region of interest, with the tip manually identified in the first frame of each video. Pixels of interest were highlighted in the binarised frame through the Shi and Tomasi minimum eigenvalue algorithm \cite{shiGoodFeaturesTrack1994}. Kanade-Lucas-Tomasi (KLT) feature-tracking algorithm initialised tracked points, with a maximum bidirectional error of 1 pixel. Points were tracked by their pixel location scaled to mm using a known reference. For 3D image analysis, two stereo cameras, fixed at 90\degree~to each other recorded film displacement from the $xz$ and $yz$ perspective, respectfully. An equivalence of the $z$ coordinates was used to determine the matrix transformation factor to scale the $yz$ coordinate positions of camera 2 to the equivalent $yz$ space of camera 1. The 3D displacement vector was then determined using the Pythagorean theorem, presented as the average displacement vector magnitude $\pm$ standard deviation (mm).

\subsubsection{DC Activation-Relaxation Response}
Characteristic relaxation and activation of the hydrated films were evaluated by applying a 5 V DC voltage for 60 seconds followed by a 60 second rest period at 0 V. Another 30 second cycle of voltage alternation at 5 V and 0 V followed the rest period. The displacement direction and magnitude in both states were recorded for at least 5 replicates \textit{via} dynamic pixel tracking.

\subsubsection{Voltage and Frequency Sweep}
The voltage dependant displacement response of hydrated films was characterised at 0, 0.5, 1, 2, 3, 5 and 7 V for 3 minutes across five replicates. 

The displacement frequency response was characterised at stimulation frequencies of 25, 50, 100, 500 and 1000 mHz for 3 minutes across triplicate samples of C, CT, CE and CTE under a 5 V square wave potential. The film response to a change frequency was then recorded, reducing the frequency from 5 Hz to 25 mHz after a 60 second activation and 60 second rest period per frequency.

Dynamic pixel tracking was used to determine the influence of input voltage and frequency on electroactive displacement.

\subsubsection{Actuation Stability Analysis}
Sustained actuation was assessed at 260, 180-second cycles over a total test time of 24 hours. Each stimulation cycle consisted of a 60 second activation phase at 5 V followed by 120 seconds of relaxation phase at 0 V. Changes in peak-to-peak displacement and time-to-peak displacement were extracted from each cycle using the displacement magnitude data from dynamic pixel tracking in a single plane ($xz$).

\subsubsection{Output pressure}
Hydrated CTE films were subjected to electromechanical actuation at 5 V for 180 seconds, following the standard protocol for DC electroactive displacement experiments. A 3 mm diameter TR150 load cell (Novatech Measurements Ltd) was placed at the tip of the film and recorded changes in applied force (N) from the tip to the load cell. The output pressure at the tip of the film actuator was calculated and is presented in Pascal (Pa) over time for five CTE replicates.

\subsection{Surface Characterisation}
\subsubsection{Scanning Electron Microscopy}
Uncoated 10 mm by 10 mm square sections from the centre of the films were imaged using a Hitachi TM4000Plus SEM at an acceleration voltage of 5 kV under vacuum at low (x100) and high (x500) magnification. Triplicates of samples from separate films were imaged to ensure representative morphologies are presented.

Electrode coated 5 mm by 20 mm films were imaged at their cross-section with an acceleration voltage of 5 kV at x2000 magnification. The film and electrode thickness were determined using ImageJ (NIH), collecting five measurements of the film thickness and three measurement of the electrode thickness per sample to calculate the mean thickness for each layer.

\subsubsection{Energy-Dispersive X-ray Spectroscopy}

Elemental compositions were obtained using energy dispersive x-ray spectroscopy with a Phenom XL G2 Desktop SEM using a silicon drift detector (Thermofisher Scientific inc.) at 15 kV and are presented as the atomic percent (at.\%) 

\subsubsection{Fourier Transform Infrared Spectroscopy}
FTIR spectra were obtained using a Shimadzu IRSpirit-QATR-S instrument over a scanning range of 4000--500 cm$^{-1}$. The peaks were fitted using the MATLAB findpeaks function to determine the most prevalent peaks between 1070--1030 cm$^{-1}$. The same function was applied to the second derivative of the curve to identify peaks neglected from the initial curve. The area under the curve was determined by integrating between the limits using the \texttt{trapz} function (trapezoidal numerical integration) of the S=O stretching region (1070--1030 cm$^{-1}$) following a baseline normalisation according to the local minima of the curve. 

\subsubsection{Charge Density}
Immobile charge density was measured using the Teorell-Meyer-Siever (TMS) theory, as established by \citeauthor{pichenyMethodsMeasurementCharge1976} in 1976 \cite{pichenyMethodsMeasurementCharge1976}. The TMS model accurately determines membrane potential for low concentration ratios using activity coefficients by \citeauthor{galamaOriginMembranePotential2016}\cite{galamaOriginMembranePotential2016}. Samples were placed between two compartments, each starting with equal 1 X PBS volumes. Sodium chloride solutions of varying concentrations were added to each compartment. Electrodes measured the potential difference related to the sample's charge density. TMS theory linked this difference to charge density, as shown in Eq. \ref{eq: Charge density voltage terms}; complete calculations are provided in the supplementary material.

\begin{equation}
\label{eq: Charge density voltage terms}
    V_0=\Delta\Phi_\text{diff}+(\Delta\Phi^{'}_{D}-\Delta\Phi^{''}_{D})-\frac{RT}{F}\ln(\frac{a^{''}_\text{Cl}}{a^{'}_\text{Cl}})
\end{equation}

\subsection{Mechanical Characterisation}
\subsubsection{Uniaxial Tensile Testing}

An Instron uniaxial tensile tester was used to assess the tensile loading mechanics on dog bone-shaped films with a gauge length of 15 mm and width of 5 mm. After sample loading, the 100 N load cell was set to zero. A constant elongation rate of 0.25 mm s$^{-1}$ was applied to the hydrated samples until failure. Force, displacement, and time were recorded in at least four replicates. The tensile modulus was calculated from the gradient of the linear portion of the stress-strain curves.

\subsection{Biological Response}
\subsubsection{Cell Culture}
Human dermal fibroblasts (HDFs, passage 12, CellsDivision (CCD-6006, Lot Number 29950)) and  mouse myoblast (C2C12s, passage 17, Sigma-Aldrich (91031101, Lot Number 22001)) were cultured in Dulbecco's modified eagle medium (DMEM, Thermo Fisher Scientific) with 10\% fetal bovine serum (FBS, Merck Life Science UK Limited) and 1\% Penicillin-Streptomycin (PS, Thermo Fisher Scientific) at 37 \degree C, 90\% RH and 5\% CO$_2$ in an incubator. The cell medium was replaced every three days and split every 7 days. Human Umbilical Vein Endothelial Cells (HUVECs, passage 8 (C2519AS)) were cultured with the EGM-2 Endothelial Cell Growth Medium-2 BulletKit (Lonza).

\subsubsection{Viability Assays}
HDFs, HUVECs, and C2C12 cells were respectively seeded at densities of 10,000, 15,000, and 20,000 cells per drop cast film sample in a 24-well plate and incubated at 37\degree C, 90\% RH, and 5\% CO$_2$. Media was changed daily. Presto Blue cell viability reagent (Life Technologies Corporation) was mixed at a 10\% concentration with fresh media at a total volume of 500 $\mu$l (50 $\mu$l of Presto Blue) and samples were left to incubate for 1 hour. Relative fluorescence (RFU) was calculated using a BMG Labtech FLUOstar Omega plate reader at a gain of 1825 for 100 $\mu$l aliquots in a transparent base NUNC 96 well plate. Cell number calculated according to calibration curves generated against known cell quantities, see supplementary Fig.11.

\subsubsection{Electrical Stimulation of HDFs}
A waveform generator (33250A, Agilent Technologies) and custom interdigitated electrodes were used to electrically stimulate \textit{in vitro} at 450 mV/mm, 1 Hz frequency, 10\% duty cycle (100 ms pulse width) for 2 hours on days 1, 3 and 5. Presto Blue\texttrademark~assays were used to determine cell populations on days 0 (3 hours post-seeding), 2, 5 and 7. Data are presented as the mean $\pm$ standard deviation of 4 replicates per sample. A schematic of this set-up is shown in supplementary Fig.12.

\subsubsection{Fluorescent Cell Staining}
Cells were fixed with 500 $\mu$l of 4\% paraformaldehyde in PBS (Thermofisher Scientific) and permeabilised by soaking in a solution of 0.5\% (v/v) Triton X-100 in 1X PBS for 15 minutes and washed three times in 1X PBS before blocking nonspecific binding sites with a filter-sterilised solution 3\% bovine serum albumin (BSA) in 1X PBS for 1 hour. Before staining with DAPI (4',6-diamidino-2-phenylindole, Thermofisher Scientific), the films were rinsed three times with a buffer of 0.1\% Tween 20 (v/v), 0.1\% filter-sterilised BSA in 1X PBS. 250 $\mu$l of a 1:10000 DAPI solution was added to each well in a 24 well plate and allowed to soak at room temperature for 10 minutes. The samples were then washed three times with 1X PBS prior to phalloidin staining. 200 $\mu$l of a 5:100 dilute Alexa Fluor® 555 Phalloidin (Cell Signaling Technology, USA) were added to each well and allowed to soak at room temperature for 15 minutes. The samples were rinsed three times after staining and left in 500 $\mu$l of 1X PBS for imaging.

The images were captured with an inverted fluorescent microscope (Eclipse Ti2, Nikon Inc, USA) at 4 $\times$ magnification. The images were processed in ImageJ, using the background subtraction feature with a rolling ball radius of 100 pixels. A Gaussian blur was then applied with a sigma of 1 pixel before using the enhance contrast function to increase pixel saturation by 0.35\%.

\begin{acknowledgments}
 This work was supported by the Royal Society of Chemistry Research Enablement Grant E20-8065 and EPSRC BIONIC Hearts New Investigator Award (EP/Y004434/1). MKB and RTM were funded by the EPSRC Doctoral Training Partnership Studentships (EP/W524311/1). VML was funded by The Rhodes Scholarship. MN acknowledges funding from an Emmanuel College Research Fellowship. We would also like to acknowledge Igor N Dyson for assistance and access to SEM and Tensile Testing, and Ben Bower for providing additional EDS spectra. 

\end{acknowledgments}

\section*{Data Availability Statement}
For the purpose of Open Access, the author has applied a CC BY public copyright licence to any Author Accepted Manuscript (AAM) version arising from this submission.
Data are available on the Oxford University Research Archive (ORA) Data Repository at 
https://doi.org/XXXXXX
.

\section{Competing interests}
MN is an inventor and MKB and CJB are contributors to patent application GB2502248.4 arising from this work.

\section{references}
\bibliography{references}

\end{document}